\documentstyle[psfig,12pt,aasms4]{article} 
\def\build#1_#2^#3{\mathrel{
\mathop{\kern0pt#1}\limits_{#2}^{#3}}}
\def\la{\mathrel{\mathpalette\fun <}}

\def\fun#1#2{\lower3.6pt\vbox{\baselineskip0pt\lineskip.9pt
        \ialign{$\mathsurround=0pt#1\hfill##\hfil$\crcr#2\crcr\sim\crcr}}}

\def\rasec{\hbox{$\,$\raise 0.6 ex \hbox{s}\kern-.35em
                  \lower 0.0 ex \hbox{.}$\,$}}        
\def\decsec{\hbox{$\,$\raise 0.0 ex \hbox{$\sec$}\kern-.45em
                  \lower 0.0 ex \hbox{.}$\,$}}         
\def\decmin{\hbox{$\,$\raise 0.0 ex \hbox{$\min$}\kern-.45em
                  \lower 0.0 ex \hbox{.}$\,$}}        
\def\gtabouteq{\,\hbox{\raise 0.5 ex \hbox{$>$}\kern-.77em 
                    \lower 0.5 ex \hbox{$\sim$}$\,$}}       
\def\ltabouteq{\,\hbox{\raise 0.5 ex \hbox{$<$}\kern-.77em 
                     \lower 0.5 ex \hbox{$\sim$}$\,$}}  

\def\be#1{\begin{equation}\label{eq:#1}}
\def\ee{\end{equation}}
\def\EC#1{(\ref{eq:#1})}
\def\bea#1{\begin{eqnarray}\label{eq:#1}}
\def\ee{\end{equation}}
\def\eea{\end{eqnarray}}

\def\msun{M_{\odot}}

\def\nh{n_{\rm HI}}
\def\nhh {n_{\rm H_2}}

\def\sec{{^\prime}{^\prime}}
\def\min{^\prime}
\def\deg{^\circ}
\def\nm{{\cal N}_{\rm min}}
\def\nb{{\cal N}_{\rm beam}}
\def\enb{\langle{\cal N}_{\rm beam}\rangle}
\def\ne{{\cal N}_{\rm eff}}
\def\ns{N_{\rm system}}
\def\rs{\rho_{\rm system}}

\newcount\notenumber 
\def\clearnotenumber{\notenumber=0} 
\def\fnote{\advance\notenumber by1 
\footnote{$^{{\the\notenumber}}$}} 
\clearnotenumber

\lefthead{Irwin, Widrow, and English} 
\righthead{HI absorption against Q0957+561}

\begin{document}

\title{An Observational Test of Dark Matter as Cold Fractal Clouds}

\author{Judith A. Irwin}
\affil{irwin@astro.queensu.ca}

\author{Lawrence M. Widrow}
\affil{widrow@astro.queensu.ca}

\and

\author{Jayanne English$^{\dagger}$}
\affil{english@astro.queensu.ca}

\vskip 0.5truein

\affil{Queen's University}
\centerline{Dept. of Physics}
\centerline{Kingston, Ontario, Canada, K7L 3N6} 
\vskip 0.5truein

\vskip 2.5truein

\noindent $\dagger$ {Now at Space Telescope Science Institute}

\begin{abstract}

Using the VLA, we have performed the first observational test of dark
matter in the form of cold, primordial fractal clouds, as envisioned
by Pfenniger, Combes, \& Martinet (1994) and Pfenniger \& Combes
(1994).  We show that, after a Hubble Time, primordial fractal clouds
will convert most of their HI to ${\rm H}_2$, but a small fraction of
HI remains which is optically thick.  This opens up a new window for
detecting dark matter which may exist in this form.  The detectability
of such gas depends on its filling factor and temperature and
therefore should be observable in absorption against a background
source, with observations of sufficient sensitivity and resolution.
The current VLA observations have made a first step towards this goal
by taking advantage of a fortuitous alignment between the extension of
the HI disk of the nearby galaxy, NGC~3079, and a background quasar,
Q~0957+561.  Our observations probe 28 independent beams against the
quasar and all of velocity space between the extension of a flat
rotation curve and a Keplerian decline for the halo region of
NGC~3079.  We do not detect any absorption features and investigate,
in detail, the implication of this result for the hypothesis that dark
matter is in the form of fractal clouds.  In particular, we calculate
the probability that our observations would have detected such clouds
as a function of the model parameters.  The chance of detection is
significant for an interesting region of parameter space
(fractal dimension $1.7\la D\la 2$ and cloud radius $30\,{\rm
pc}<R_c<3\,{\rm kpc}$) and rises above 95\% for a
small region of parameter space.  While our analysis does not rule out
fractal clouds as dark matter, it does lay out the groundwork for
future, more sensitive observations and we consider what form these
might take to probe the range of possible cloud properties more
deeply.  It is interesting that the observations can rule out cold
optically thin HI gas, if it exists, to a limit of 0.001\% of the dark
matter.  In contrast, the existence of cold HI in a fractal hierarchy
would be an efficient way of hiding dark matter.

\end{abstract}

\keywords{cosmology: dark matter --- galaxies: halos, individual (NGC 3079)
--- quasars: absorption lines --- ISM: clouds}

\section{Introduction}
\label{intro}

The question of extended disks around spiral galaxies, heavy halos,
and indeed, the very nature of dark matter, has engaged astronomers
for many years and dates to at least the early observations of flat
rotation curves (e.g. \cite{rub80}).  Since that time, observations
too numerous to list have been carried out in attempts to search for
evidence, direct or indirect, for this dark matter and to constrain
its nature.  Whatever the composition, there is now strong evidence
that the dark halos surrounding spiral galaxies can extend to large
distances.  According to Ashman (1992), there is general agreement
that the dark halo of the Milky Way, for example, extends to a radius
between 100 kpc and 200 kpc.

It has been known for some time that primordial nucleosynthesis favors
a density in baryons that is significantly greater than the density in
luminous matter (e.g., Copi, Schramm, \& Turner 1995).  While
recently, some of the results of primordial nucleosynthesis have been
called into question, the conclusion that there are nonluminous
baryons in the Universe remains firm and is in fact supported by other
arguments from astrophysics and cosmology (Steigman, Hata, \& Felten
1999).  The question remains whether baryonic dark matter contributes
significantly to halos in spiral galaxies.  Baryonic dark matter has
attracted considerable attention of late (see Carr \cite{car94}),
especially in the light of recent reports of gravitational
microlensing events toward the Large and Small Magellanic clouds,
presumably (though not conclusively) due to massive compact halo
objects (MACHOs) in the Milky Way (Alcock et al.\,1997a, \,1997b).  At
present, microlensing experiments are unable to fix the MACHO fraction
in the Galaxy.  In addition, the most probable mass for the lenses
($\sim 0.5-1 \msun$) presents a puzzle since this rules out brown
dwarfs and giant planets, initially considered to be the most
plausible candidates for MACHOs.

A number of authors have considered cold gas as a dark matter
candidate.  Pfenniger, Combes, \& Martinet (1994) and Pfenniger \&
Combes (1994) (hereafter PCM and PC) have argued that the flat
rotation curves of spiral galaxies might be accounted for entirely by
cold, dense H$_2$ gas clouds.  Henriksen \& Widrow (1995) and Draine
(1998) have suggested that very dense clouds might themselves act as
gravitational lenses while Gerhard \& Silk (1993, 1996) and de Paolis
et al.\, (1995a, 1995b) have considered the possibility that gas might
coexist with a population of MACHOs.

PCM and PC, in particular, have proposed that dark matter consists of
cold primordial gas clouds in extended disks or halos around spiral
galaxies where the gas within an individual cloud is distributed in a
fractal structure.  Far from sources of heating, the gas is
thermalized to the 3 K background, assisted by small quantities of
${\rm H}_2$ ice.  PCM suggest that the slow accretion of gas toward
the visible disk solves the ``gas consumption problem" (i.e. that the
timescale for gas consumption, as implied by current star formation
rates, is shorter than the age of the galaxy) as well as the
``disk-halo conspiracy" (i.e. the smooth, flat rotation curves through
the transition region from disk to halo dominated rotation).  Upon
re-evaluation of the stellar recycling of gas, the former problem may
not be as serious as previously thought (Kennicutt et al. 1994), but
the latter problem remains.  The PCM and PC proposal is motivated by
fractal interpretations of the hierarchical structures observed over a
wide range of scales for atomic and molecular interstellar clouds
(see, e.g. Diamond et al. 1989, Vogelaar \& Wakker 1994, and Elmegreen
1996).  The HI in 7 members of the M~81 group (Westpfahl et al. 1999)
and possibly in the Small Magellanic Cloud (SMC) (Stanimirovic et
al. 1999) have also recently been shown to have a fractal structure.
PCM suggest that the gas is in the form of 3 K molecular hydrogen
which is unobservable directly and, if primordial, would also be
unobservable via tracers such as CO.  However, if a component of
atomic hydrogen is present (see \S~\ref{composition}), it could be
detected in absorption against a background continuum source with
observations of sufficient resolution and sensitivity.

In this paper we describe the first observational test of the PCM and
PC model.  Specifically, we search for HI absorption associated with a
foreground galaxy, NGC~3079, against the background quasar,
Q~0957+561, which is fortuitously located nearby.  Some preliminary
results from this work have been presented in Irwin et al.\,(1999).
In \S\ref{fractal_clouds}, we describe the PCM and PC model and
discuss the neutral hydrogen content of the gas.  In
\S\ref{observations_data} we describe NGC~3079/Q~0057+561, the
observations and data reduction, and in \S\ref{results} we present the
results.  No significant absorption features are detected.
The observations are extremely sensitive ($N({\rm
HI})=1.4\times 10^{17}\,{\rm cm}^2$) and indicate that no more than
$0.001\%$ of the dark matter halo can be in the form of cold {\it
diffuse} HI.  However, as we see in \S\ref{constraints}, we are unable
to rule out fractal clouds as a dark matter candidate since the
probability of detection is high ($>95\%$) over a limited region of
parameter space.  This particular experiment lays the groundwork for
future, more sensitive observations, as discussed in
\S\ref{discussion}.

\section{The Fractal Cloud Model}
\label{fractal_clouds}

\subsection{Elementary Cloudlets}
\label{cloudlets}

The elementary building blocks in the model of PCM and PC are
Jupiter-mass objects called ``cloudlets" (``clumpuscules" by PC) which
describe a natural minimum mass for fragmentation.  This scale emerges
by setting the free-fall time equal to the Kelvin-Helmholtz time and
assuming virialization.

The cloudlet parameters, i.e. mass, radius, volume density, column
density, and thermal line width, respectively, are given by:

\be{massletgen}
M_* \simeq 4.0 \times 10^{-3}~T^{1/4}\,\mu^{-9/4}\,f^{-1/2} ~~~ \msun
\ee
\be{radletgen}
R_*\simeq 1.5 \times 10^{2}~T^{-3/4}\,\mu^{-5/4}\,f^{-1/2}~~~ {\rm AU}
\ee
\be{densletgen}
n_*\simeq 1.1 \times 10^{8}~T^{5/2}\,\mu^{3/2}\,f~~~ 
{\rm H~cm^{-3}}
\ee
\be{surfletgen}
N_* \simeq 3.2 \times 10^{23}~T^{7/4}\,\mu^{1/4}\,f^{1/2}~~~ 
{\rm H~cm^{-2}}
\ee
\be{velocitylet}
v_* \simeq 9.1 \times 10^{-2}~T^{1/2}\mu^{-1/2}~~~~{\rm km ~s^{-1}}
\ee

\noindent where $T$ is the temperature, $\mu$ is the mean molecular
weight, and $f$ is a factor which accounts for departures from
spherical symmetry and blackbody radiation (PC).  
Note that $n_*$ is the mass
density divided by the hydrogen mass, $m_H$ 
({\it not} by $\mu\,m_H$), i.e. $n_*\equiv 3M_*/4\pi
R_*^3 m_H$, and therefore includes all forms of hydrogen as well as
helium.  The same holds for the surface density, $N_*$.  
These equations imply that warmer cloudlets are smaller, more massive, and
thus more stellar-like.

For $T\,=\,3\,$K, $0.1\,\le\,f\,\le\,1$ (see PC)
and $1.3\,\le\,\mu\,\le 2.3$ (primordial neutral gas), 
the resulting cloudlet parameters are,
$M_*\,=\,0.8\,$--$\,9.2 \times 10^{-3}~M_\odot$,
$R_*\,=\,23\,$--$\,150~{\rm AU}$, 
$n_*\,=\,0.25\,$--$\,6 \times 10^9~{\rm cm^{-3}}$,
$N_*\,=\,0.73\,$--$\,2.7 \times 10^{24}~ {\rm cm^{-2}}$, and
$v_*\,=\,0.10\,$--$\,0.14~ {\rm km ~s^{-1}}$.
Taking $\mu\,=\,2.3$ (neutral H$_2$ $+$ He,
see \S~\ref{composition}) and adopting
$f\,=\,1$, we have

\be{masslet1}
M_* = 0.81 \times 10^{-3} ~~~\msun
\ee
\be{radlet1}
R_* = 23 ~~~ {\rm AU}
\ee
\be{denslet1}
n_* = 6.0  \times 10^{9}~~~ {\rm H~cm^{-3}}
\ee
\be{surflet1}
N_* = 2.7 \times 10^{24}~~~ {\rm H~cm^{-2}}
\ee
\be{vellet1}
v_* = 0.10 ~~~{\rm km~s^{-1}}
\ee

\subsection{Composition of the Cloudlets}
\label{composition}

Cloudlets at such high densities, if they exist, should consist only
of neutral gas.  While low density galactic HI disks are expected to
be truncated by the extragalactic ionizing radiation field (Corbelli
\& Salpeter 1993, Maloney 1994, Dove \& Shull 1994, see also
Bland-Hawthorn et al. 1997), a simple application of the
photoionization equilibrium equation shows that the background UV
photon rate at the current epoch (4 $\times$ 10$^{-14}$ s$^{-1}$;
Haardt \& Madau 1996) is insufficient to produce any significant
ionization in such high density cloudlets.

True primordial gas consists of atomic hydrogen since the
``freeze-out" fraction of molecular hydrogen is only [H$_2$/H] = 1.1
$\times$ 10$^{-6}$ (Galli \& Palla 1998) and dust is not available to
assist in the conversion from HI to H$_2$.  However, for these high
density cloudlets, the 3-body reactions

\be{r1}
{\rm H + H + H \rightarrow H_2 + H}
\ee

\be{r2}
{\rm H + H + H_2\rightarrow H_2 + H_2}
\ee

\noindent may be important in converting HI to H$_2$ (see Palla\, et
al. 1983).  If we consider only these reactions, then the evolution of
the number densities $\nh$ and $\nhh$ are governed by the
equations:

\be{nheq1}
\frac{d\nh}{dt} = -2{\cal R}_1\nh^3-2{\cal R}_2 \nh^2 \nhh
\ee

\be{nH2}
\frac{d\nhh}{dt} = {\cal R}_1\nh^3+{\cal R}_2 \nh^2 \nhh
\ee

\noindent where ${\cal R}_1$ and ${\cal R}_2$ are the rate coefficients 
for reactions \EC{r1} and \EC{r2},
respectively.  These have not been measured
at the temperatures and densities being considered here.  Cohen and
Westberg (1983) review the available data and recommend values, ${\cal
R}_1\,=\, 8.8\,\times\,10^{-33}$ cm$^6$ s$^{-1}$, independent of
temperature, $T$, and ${\cal R}_2\,=\, 2.8\,\times\, 10^{-31}\,
(T/K)^{-0.6}$ cm$^6$ s$^{-1}$ above 80 K.  In the absence of other data,
we assume the above temperature dependencies hold down to 3 K so that
for the case at hand, ${\cal R}_1$ is as given above and ${\cal
R}_2=1.4\times 10^{-31}\,{\rm cm^6\, s^{-1}}$.

When the cloudlets first form (time $t_0$), the gas is almost entirely
in atomic form.  However, on a timescale of order $(\nh^2(t_0){\cal
R}_1)^{-1}\sim 10^5\,{\rm yrs}$, most of the hydrogen is converted to
H$_2$.  Thereafter, the second reaction dominates and we have
$\nhh\simeq \nh(t_0)/2$ and

\be{HIlatetimes}
\nh(t) \sim \frac{1}{{\cal R}_2 \,\nh(t_0)\,
\left (t-t_0\right )}
\ee

\noindent For $\nh(t_0)$ = $0.25 \,- \,6 \,\times \,10^9$ cm$^{-3}$
(\S~\ref{cloudlets}) and ($t\,-\,t_0$) = 10 Gyr, Eqn.~\EC{HIlatetimes}
yields $\nh(t)$ = $4.0\,\times\,10^3$ - $9.5\,\times\,10^4$
cm$^{-3}$.  We take $\nh(t)$ to represent the present density of HI
in a cloudlet, i.e. $n(HI)_*$. This implies that the fraction of the
cloudlet in the form of HI, $f(HI)\,=\,n(HI)_*/n_*$, is small, i.e.
$f(HI)\,=\,7\,\times\,10^{-7}\, - \,2\,\times\,10^{-5}$.

If HI coexists with H$_2$ throughout the cloudlet, then the HI column
density will be $N(HI)_*\, = \,2\,\nh(t)\,R_*\,=\,
2.8\,\times\,10^{18}\,-\, 4.3 \,\times\,10^{20}$ cm$^{-2}$.
Since in general,

\be{nheq2}
\frac {N(HI)}{[\rm cm^{-2}]} = 1.82 \times 10^{18} \int \tau_v~ 
\frac{T_s}{[\rm K]}~ \frac{dv}{[\rm km s^{-1}]}
\ee
where $\tau_v$ is the HI optical depth at velocity, $v$, and $T_s$ is
the spin temperature (kinetic temperature) of the gas, the above
parameters imply that $\tau \,\sim 5 \, - \, 700$, and hence even this
small fraction of HI in the cloudlets would be optically thick.  The
detectability of such gas depends not on quantity, but on 
temperature and filling factor.   Thus, even though HI may constitute
a small fraction of the mass of the cloudlet, it can still provide
a sensitive observational probe of the gas, if it exists.

\subsection{Fractal Clouds}
\label{clouds}

Cloudlets coalesce hierarchically in a fractal distribution to form
larger structures called clouds (denoted with subscript, $c$,
below). The clouds have parameters related to the parameters of the
cloudlets via,

\be{masscloud}
M_c = {\cal N}_c\,M_*\,=\,\left (\frac{R_c}{R_*}\right )^D M_*
\ee
\be{denscloud}
n_c = \left (\frac{R_c}{R_*}\right )^{D-3} n_*
\ee
\be{surfcloud}
N_c  = \left (\frac{R_c}{R_*}\right )^{D-2} N_*
\ee
\be{velcloud}
v_c = \left ( \frac{R_c}{R_*}\right)^{(D-1)/2}v_*
\ee

\noindent where ${\cal N}_c$ is the number of cloudlets in a cloud,
$v_c$ is the velocity dispersion of the cloud, $R_c/R_*$ is the
dynamic range of the fractal structure, and $D$ is the {\it fractal
dimension}.  A pure diffuse smooth medium has $D=3$ and lower values
of $D$ correspond to progressively lower filling factors.  In general,
we expect $D\,\ltabouteq\,2$ since for $D\,>\,2$, the clouds tend to
dissolve due to internal collisions (PC). Molecular clouds in the ISM
exhibit fractal structure with $1.4\,\le\,D\,\le\,2$.  Recent results
for HI in the M~81 group of galaxies give $1.2\,\le\,D\,\le\,1.5$
(Westpfahl et al. 1999) and for HI in the SMC, $D\,=\,1.5$
(Stanimirovic et al. 1999).

\section{Observations and Data Reduction}
\label{observations_data}

\subsection{The NGC~3079 - Q~0957+561 Pair}
\label{galaxy-qso pair}

A fortuitous alignment of the extension of the major axis of the
galaxy, NGC~3079, with the background quasar, Q0957+561 has provided
the opportunity to search for cold gas in an extended halo or an
extended disk around the foreground galaxy (see Figure 1).  NGC~3079 is
at a distance of D = 15.6 Mpc (H$_0$ = 75 km s$^{-1}$ Mpc$^{-1}$;
$z~=~0.004$) and has been mapped in HI by Irwin \& Seaquist (1991) and
Irwin et al. (1987). The outermost point at which emission is observed
is denoted by a small star in Fig. 1.  Q0957+561 is a gravitationally
lensed system (\cite{wal79}) at a redshift of z = 1.4 and has also
been extensively studied (e.g. Schmidt \& Wambsganss 1998 and
references therein).  The projected separation between the center of
NGC~3079 (taken to be the kinematic center given by Irwin \& Seaquist
1991), and Q0957+561 (taken to be the position of component A of RA
(J2000) = 10$^{\rm h}$ 01$^{\rm m}$ 20$\rasec$69, DEC (J2000) =
55$\deg$ 53$\min$ 55$\decsec$9) is 14.13 arcminutes, or 64.1
kpc at the distance of NGC~3079.  For our purposes, we consider the
lensed quasar to be simply a source of background radio continuum
emission.

\subsection{The Observations}
\label{observations}

Observations were carried out in the 21 cm spectral line on 03 January
1997 in the A configuration of the Very Large Array (VLA)\footnote
{\small The National Radio Astronomy Observatory is a facility of the
National Science Foundation operated under cooperative agreement by
Associated Universities, Inc.}.  The flux calibrators were J1331+305
(3C286) and J0137+331 (3C48) which had flux densities at the central
observing frequency of 14.79 Jy and 16.04 Jy, respectively, using the
latest VLA calibration scale.  The phase calibrator, J0957+553, for
which we find a flux density of 2.45 $\pm$ 0.01 Jy, is separated from
the source by only 43 arcminutes and was observed every $\sim$ 20
minutes.  All calibrators had UV restrictions which were applied
during calibration.  The field center was placed 13 arcseconds south
of the quasar to avoid baseline-dependent map errors which tend to
show up at the field center (Ekers 1989) and, for highest sensitivity,
the 4IF mode was used.
 
The systemic velocity of NGC~3079 is 1117 km s$^{-1}$ (heliocentric,
optical definition) and the peak of the rotation curve is offset
$\pm$215 from this value (Irwin \& Seaquist 1991) with the
background quasar occurring on the blueshifted side of the galaxy.
Thus the central observing velocity was set to 974 km s$^{-1}$, which
is between the velocity of a flat rotation curve for NGC~3079 extended
to the projected distance of the quasar (i.e. 902 km s$^{-1}$) and the
velocity expected at the position of the quasar for a Keplerian
fall-off (1010 km s$^{-1}$).  The bandwidth of 163.4 km s$^{-1}$
encompasses both of these possibilities (though some channels at the
band edges were lost due to a loss of sensitivity).
At these frequencies, there should be no contamination by Galactic HI
or by the lensing galaxy (z = 0.36) or known absorption systems (z =
1.39 and z = 1.12, \cite{wal79}) along the line of sight.

With on-line Hanning smoothing and 63 channels, the final channel
width (= the spectral resolution) was 2.593 km s$^{-1}$.  The primary
flux density calibrators were also used as bandpass calibrators and
were observed 4 times during the 10 hours of observations.  The
calibration observation closest in time to the source observation was
applied.  Examination of the average bandpass for each time showed
that the maximum temporal change in the bandpass over the 4 scans was
of order 0.2\%.

UV data for each channel were Fourier Transformed and cleaned
(e.g. Cornwell \& Braun 1989) using uniform weighting (e.g.  Sramek \&
Schwab 1989) 
(RA-DEC-Velocity) cube.  A uniform weight clean continuum
image was also made by averaging together the UV data for all channels
(except the noisy ones at the band edge) before Fourier
Transformation.  The data were phase-only and then phase+amplitude
self-calibrated (e.g. Cornwell \& Fomalont 1989) until no improvement
in the rms resulted.  The final rms noise per channel is 0.61 $\pm$
0.01 mJy beam$^{-1}$, where the error represents the 1$\sigma$
variation over all channels.  This is $\sim$ 1.3 times the theoretical
rms for this weighting.

A continuum-subtracted cube was made by subtracting a linear fit to
the visibilities over low-noise channels from each total emission
channel.  The resulting rms noise in the continuum-subtracted maps is
0.58 $\pm$ 0.01 mJy beam$^{-1}$.  Channel-to-channel variations due to
the bandpass calibration are smaller than this.  The spectral dynamic
range, i.e. the ratio of peak continuum intensity to rms noise in the
channel maps is $\sim$ 300/1.

Naturally weighted cubes and continuum images were also made (with and
without a UV taper), but since the rms was not significantly improved
and the beam size was larger, subsequent analysis proceeded on the
uniform weighting maps.

A summary of the observing and map parameters is given in Table~1.

\section{Results}
\label{results}

The continuum map is shown in Figure 2, with components labelled
according to Greenfield et al. (1985) and Avruch et al. (1997).  The
lensing galaxy is associated with radio source, G, which is visible as
a northern extension to component B.  This map is very similar to the
18 cm image obtained by Avruch et al. (1997) from archival VLA data
and confirms their discovery of sources R1, R2, and the northern
source, N.  Although not visible in Fig. 2, we also detect their
southern source, S, in the naturally weighted map.

The continuum-subtracted cube was inspected for signs of emission or
absorption features.  No HI features were detected.  In Fig. 3, we
show spectra taken at the positions of the peaks of continuum
components A, B, C, D, and E.  The last panel shows the total flux as
a function of velocity, integrated over a region in which the
continuum emission is greater than 10 $\times$ the continuum map rms
noise value.  These show no evidence for features above $\sim$ 3
$\times$ the rms noise level per channel.  Note that the noise in
these spectra is dominated by the rms map noise, rather than by
variations in the bandpass calibration which are smaller
(\S~\ref{observations}).  There is a slight hint of an absorption
feature near zero velocity for components A, C, and E.  However, this
feature disappears if a spectrum is taken several pixels away from the
continuum peaks but still within the same beam, and it also disappears
if an average spectrum is taken over the beam at these positions.
Moreover, if a spectrum is taken at a random position off-source,
similarly sized features result.  Thus, these features are only noise
peaks.

We also formed 0th and 1st moment maps, i.e. total intensity and
intensity-weighted mean velocity maps, respectively (not shown) in
which low intensity emission was cut-off so that the higher intensity
peaks would be highlighted.  Again, there is no evidence for emission
or absorption in any channel.

In the event that HI exists at a low level over many channels,
we also created a continuum-subtracted cube by doing linear fits
to the visibilities over channels near the 
(low-noise) ends of the useable band,
only.  No evidence for any HI features could be seen in this cube either.

\section{Constraints on Fractal HI in a Dark Matter Halo}
\label{constraints}

\subsection{Density of HI within NGC 3079}
\label{density_at_quasar}

Our ability to detect clouds of the type proposed by PCM and PC
depends both on the structure of individual clouds (the parameters
$R_c$, $R_*$, and $D$) and the overall mass density in clouds.  To
quantify the latter, consider the hydrogen clouds in NGC 3079 to be
smoothed into a continuous mass distribution throughout the halo of
the galaxy.  The line of sight to the background QSO intersects
various galactocentric radii of this mass distribution.  We define the
hydrogen mass density at the radius closest to the center of NGC 3079
along this line of sight to be $\rs$.

Flat rotation curves observed in the outer parts of spiral galaxies
suggest that dark matter halos follow an $r^{-2}$ density law.  Beyond
this, little is known about the distribution of dark matter (e.g.,
shape of the halo).  For definiteness, we assume an isothermal
spheroid model for the halo of NGC 3079:

\be{density}
\rho_{\rm DM} = \frac{\lambda(q) v_c^2}{4\pi G }\frac{1}{R^2 + z^2/q^2}
\ee

\noindent where $v_c$ is the circular rotation speed, $R$ is the
galactocentric radius in the plane of the disk, $z$ is the coordinate
perpendicular to the plane ($r^2\,=\,R^2\,+\,z^2$), $q$ is the
flattening parameter ($q<1$ for an oblate halo), and
$\lambda(q)=\sqrt{1-q^2}/(q\, {\rm arccos}\,q)$ is a geometric factor
equal to $1$ for $q=1$ and rising to $6.8$ for $q=0.1$.  We have
ignored the possibility of a core radius since we are only interested
in the outer parts of the halo.

Roughly speaking, a maximum value $\rho_{\rm max}$ for $\rs$ is
obtained by assuming that the entire halo is composed of clouds (i.e.,
$\rs=\rho_{\rm DM}$), that the halo is highly flattened ($q=0.1$) and
that the projected position of the quasar is in the equatorial plane
of the halo ($R=64 \,{\rm kpc};~z=0$).  Taking $v_c$ = 215 km s$^{-1}$
(Irwin \& Seaquist 1991) we find $\rho_{\rm max}=9.6\times
10^{-26}{\rm g\,cm^{-3}}$.  Let $F_H\equiv\rs/\rho_{\rm max}$ be the
mean density in clouds relative to $\rho_{\rm max}$.  $F_H=1$
therefore corresponds to the ``best-case'' scenario for detecting
clouds.  Any possible effect which might reduce $\rs$ is parametrized
by $F_H$.  

Integrating Eq.\,\EC{density} from -$\infty$ to $\infty$ along the
line of sight to the QSO and expressing the result in terms of a
maximum surface density $N_{\rm max}$ (i.e., the surface density
assuming $q=0.1$ and $z=0$) and the parameter, $F_H$, yields the
following for the surface density:

\be{surfdens}
\ns \,= \,
N_{\rm max} \, F_H \,=\,
\frac{\lambda(q)\,{v_c}^2}{4G m_H R}\,F_H\,=\,
3.6 \times 10^{22} \,F_H\, ~{\rm H\,cm^{-2}}
\ee

\noindent
The units, ${\rm H\,cm^{-2}}$, are the same as in
Eq.\,\EC{surfletgen} and therefore represent the 
mass per unit surface
area normalized by $m_H$,
 rather than the number of molecules per unit surface area.
In deriving this expression, we have assumed that
the density law is $r^{-2}$ well beyond $R=64\,{\rm kpc}$.  Obviously,
one could introduce an additional parameter to allow for the
possibility that the density law is steeper than $r^{-2}$ in the outer
parts of the halo.

\subsection{Observational Parameters and Their Relation to Fractal
Parameters}
\label{observational-fractal parameters}

 \subsubsection{The Observational Parameters}
\label{obs_parms}

For HI absorbing gas in front of a uniform background continuum
source, the measured brightness temperature, $T_B$, in a given beam
and velocity channel (i.e. a given resolution element) is given by
 
\be{Tb}
T_{B}\,=\,\left [T_c\,e^{-\tau}\,+\,T_s(1\,-\,e^{-\tau})\right ]
\,{\cal N}\,f_b\,f_v\, +
\,T_c(1\,-\,{\cal N}\,f_b\, f_v)
\ee
where $T_c$ is the brightness temperature of the background continuum,
$T_s$, $\tau$, are the spin temperature and optical depth,
respectively, of the gas, $f_b$, $f_v$ are the filling factors for the
beam area and velocity for a single cloud, and $\cal N$ is the number
of clouds in the resolution element.  Thus, the quantity, ${\cal
N}\,f_b\,f_v$ is the filling factor for a resolution element.  The
factors, $f_b$ and $f_v$, are dealt with separately since there are
different conditions for saturating the beam and channel, respecively.
The above equation reduces to the usual one for HI if this
filling factor is unity.  Aside from a constant of proportionality
given by the Rayleigh-Jeans relation (T/I, Table 1), $T_{B}$ is
represented by the total emission channel maps.

After continuum subtraction,

\be{hionly1}
\Delta\,T_{B}\,=\,(T_s\,-\,T_c) (1\,-\,e^{-\tau})\,{\cal N}\,f_b\,f_v
\ee

\noindent where $\Delta T_B$ represents the measured quantity in the
continuum-subtracted cubes.

\noindent Since for most emission, $T_c\,>>\,T_s$, then 
we can form the ratio,

\be{hionly}
\frac{\Delta T_B}{-T_c} = \left\{ \begin{array}{ll}
{\cal N}\,f_v\,f_b & \mbox{for $\tau > 1$ (fractal clouds)} \\
\tau\,{\cal N}\,f_v\,f_b & \mbox{for $\tau < 1$ (optically thin gas)}
		\end{array}
	\right. 
\ee
 
\noindent Figure 4 shows a map of the quantity
$a\equiv\Delta\,T_{B}/{(-T_c})$ for a single velocity channel where we
have taken $\Delta\,T_{B}$ to be the negative of the 
rms map noise.  To ensure that
$T_c\,>>\,T_s$ for all points, we have cut off all continuum emission
less than 1.5 mJy beam$^{-1}$ = 583 K.  Since the rms noise is
constant over the map, this map has the appearance of the reciprocal
of the continuum map.  Such a map could be formed for any velocity
channel, but the map for any other channel would be identical, since
both the continuum distribution and rms noise are the same, channel to
channel.  Note that this ratio map would be a map of upper limits to
the optical depth under the more common assumptions of unity filling
factor and optically thin gas.

Figure 4 provides observational constraints on the filling factor of
optically thick fractal clouds through Eqn.~\EC{hionly}.  If $a$ is
the value read from Figure 4 for a particular resolution element, then
our non-detection implies

\be{condition_1}
{\cal N}\,f_b\,f_v\,<\,3\,a
\ee

\noindent where we impose a 3$\sigma$ upper limit,
i.e. [$3\,a$], on detectability.

The minimum of Figure 4 has a value 0.0033.  Thus the strongest
constraint we have for a single beam is $3a= 0.01$,
i.e. less than 1\% of this beam/velocity resolution element
is filled with cold, fractal clouds.   Of course, the
clouds are distributed randomly and so we can tighten our constraint
by considering the full data set.  Our observations cover
approximately 28 independent beams
in any given channel and there are 
163 channels\footnote{We have taken the full velocity coverage in
this analysis, rather than the slightly reduced coverage due to
the presence of a few noisy end channels.} 
or 4564 independent beam/velocity resolution elements, in total.

\subsubsection{Model Parameters}

In this section, we apply the observational constraints to a family of
fractal models defined by the parameters $R_c/R_*,~ D,$ and $F_H$.  We
assume that the cloudlets are optically thick in HI
(\S\ref{composition}) and that their physical characteristics are
given by Eqs.\,\EC{masslet1}-\EC{vellet1}.  We focus attention on
values for $D$ between $1$ and $2$ (\S\ref{clouds}) and values for
$R_c/R_*$ less than $3\times 10^7$ which 
corresponds to a cloud radius $\simeq 3\,{\rm kpc}$.  It is difficult
to imagine a cloud $64\,{\rm kpc}$ from the center of the galaxy with
a size much bigger than this.  Indeed, such a cloud would subtend
2.7$\deg$ if placed around the Milky Way, albeit with low covering
factor.

For each point in parameter space, we calculate the probability that a
cloud would have been detected.  Recall that for a
positive detection in a given resolution element, we require ${\cal
N}\,f_b\,f_v\,>3a$.  It is straightforward to determine $f_b$ and
$f_v$ as a function of $R_c/R_*,~ D,$ and $F_H$ (see below) while
${\cal N}$ can only be determined in a statistical sense.  To see how
this works, consider first a single beam.  Let $P(\nb)$ be the
probability of finding $\nb$ clouds in this beam and $P(\nb|\ne)$ be
the conditional probability that at least one of its channels contain
$\ne$ or more clouds, given $\nb$.  By Bayes's theorem, the probability
that at least one channel will contain $\ne$ or more clouds is

\be{bayes}
P(\ne)=P(\nb)P(\nb|\ne)
\ee

The probability of having a detection in this beam is therefore

\be{prob1}
P=\sum_{\ne=\nm}^\infty P(\ne)
\ee

\noindent where 

\be{nmin}
\nm=\frac{3a}{f_b\,f_v}
\ee

\noindent is the minimum number of clouds required for a detection.  

It is straightforward to extend this analysis to all 28 beams.  We use
$j=$1--28 to label the different beams, i.e., $P$, $a$, and $\nm$ all
carry the subscript $j$.  The probability for detecting at least one
cloud within the entire data set is
\be{totalp}
{\cal P}=1-\prod_j \left ( 1-P_j\right )
\ee
The calculation of ${\cal P}$ requires, for each
point in parameter space, $f_v$, $f_b$, $P(\nb)$, and $P(\nb|\ne)$.
We determine these now, except for $P(\nb|\ne)$ which is presented
in the appendix.

Recall that the velocity dispersion of a cloud is $v_c$ given by
Eq.\,\EC{velcloud} while the width of a velocity channel is $\Delta
v\equiv 2.59\,{\rm km s^{-1}}$.  A contour plot of $v_c$ as a function
of $R_c/R_*$ and $D$ is given in Figure 5.  We estimate $f_v$ by assuming
that for $v_c\ge\Delta v$, the cloud covers the entire channel while
for $v_c<\Delta v$ the cloud covers a fraction $v_c/\Delta v$ of the
channel: 
\be{velocityff} 
f_v={\rm min}\left\{1,~v_c/\Delta v\right\}
\ee 
This ignores the obvious complication that a cloud can straddle
more than one velocity channel.  In particular, for $v_c\simeq \Delta
v$, a cloud will generally cover some fraction of two neighboring
channels.

The beam filling factor, $f_b$, depends on both the size of the cloud
relative to the beam, and the area filling factor for the cloudlets in
a cloud.  If the cloud is larger than the beam ($R_c>R_b$ where
$R_b=47\,{\rm pc}$ is the mean beam radius) then $f_b=\left
(R_b/R_*\right )^{D-2}$, in analogy with Eq.\,\EC{surfcloud}.
Conversely, if $R_c<R_b$ then 
$f_b=\left ({A_c/A_b}\right )\left
(N_c/N_*\right )= \left (R_c/R_b\right )^{2}\left (R_c/R_*\right
)^{D-2}$ where $A_c$ and $A_b$ are the areas of the cloud and beam
respectively.  These results can be summarized in the following
expression: 
\be{beamff} 
f_b=\left (\frac{R_{\rm min}}{R_b}\right )^2
\left (\frac{R_{\rm min}}{R_*}\right )^{D-2}\nonumber 
\ee 
where $R_{\rm min}={\rm min}\left (R_b,\,R_c\right )$.  From
Eqs.\,\EC{nmin},\EC{velocityff} and \EC{beamff} we construct a contour
plot (Figure 6) for $\nm$ assuming $a=0.0033$ (our most sensitive
beam).  From this figure, we can read off the number of clouds
required in our most sensitive beam for a positive detection.  For
example, if the cloud system is described by the parameters $D=1.64$
and $R_c/R_*=10^6$, then we would require at least one cloud in this
beam for a positive detection.  Likewise, for $D=1.47$ and
$R_c/R_*=10^6$, we would require about 10 clouds in this one beam.

We now calculate $\enb$, the expectation value for the number of
clouds in a single beam.  To do this, consider an area $A$ large
enough to contain a representative sample of clouds.  The expected
number of clouds in $A$ will be ${\cal N}_A=\left (\ns / N_c\right )
\left (A/A_c\right )$.  Next consider a single beam in $A$.  If
$R_c>R_b$ there are three possibilities: cloud and beam do not
intersect; cloud partially covers beam; cloud covers the entire beam.
Likewise, the three possibilities for $R_c<R_b$ are: cloud and beam do
not intersect; part of cloud covers beam; cloud entirely within beam.
To avoid a rather awkward treatment that would attempt to take into
account each of these possibilities, we use, as criteria for the cloud
to be ``in'' the beam, that either the center of the beam is within
the cloud perimeter (the $R_c>R_b$ case) or the center of the cloud is
within the beam perimeter (the $R_c<R_b$ case).  The probability for a
single cloud in $A$ to be also in the beam is therefore $A_{\rm
max}/A$ where $A_{\rm max}\equiv {\rm max}(A_b,A_c)$.  The
probability that none of the clouds in $A$ overlap with the beam is
$\left (1- A_{\rm max}/A\right )^{{\cal N}_A}$ $\approx$ $\left
(1-{\cal N}_A\,A_{\rm max}/A\right )$.  Since $A\ge {\cal
N}_A\, A_{\rm max}$, the expected number of clouds in the beam is then,

\be{nbeam}
\enb ~\simeq~
{\cal N}_A\,A_{\rm max}/A
~=~
\frac{\ns}{N_c}\left (\frac{A_{\rm max}}{A_c}\right )
~=~\frac{F_H N_{\rm max}}{N_*}
\left(\frac{R_c}{R_*}\right )^{-D}
\left (\frac{R_{\rm max}}{R_*}\right )^2
\ee
where $R_{\rm max}\equiv {\rm max}\left (R_c,R_*\right )$.

\noindent A contour plot of $\enb$ is given in Figure 7.  We have
assumed that $F_H\,=\,1$, but the contours can be shifted proportionately
for lower $F_H$ according to Eqn~\EC{nbeam}.  Note, however, that
$F_H$ approaching 1 is the only astrophysically interesting case.

The desired probability function, $P(\nb)$ is given by the Poisson
distribution:
\be{poisson}
P(\nb)= \frac{\enb^{\nb} e^{-\enb}}{\nb!}
\ee
The calculation of $P(\nb|\ne)$ is  more involved and 
therefore left to the appendix.

All of the ingredients are now in place and we can perform the
calculation for ${\cal P}$, the probability for detecting at least one
cloud in the entire dataset.  Recall that an oblate halo filled with
clouds, i.e. $F_H=1$, is the best case scenario for detection.  This
is shown in Figure 8.  The consequences of the beam area filling
factor, Eq.\,\EC{beamff}, are evident in the plot.  For example,
clouds with $D<1.7$ have $f_b\la 0.02$ even if they fill the beam
($R_c>R_b$).  Likewise, clouds with $R_c < 0.1R_b\simeq 10^{4.5}R_*$
have $f_b<0.01$ even for $D=2$.  In these regions of parameter space
one requires two or more clouds in a given resolution element which is
unlikely.  If fractal clouds exist in this range of parameter space,
our observations would not have been able to detect them.

Our strongest constraints apply to the region,
$1.75\la D \la 1.85$
  and
$100\,{\rm pc}\la R_c\la 3\,{\rm kpc}$ 
($R_c > R_b$) for which we find the highest
probability, i.e.  $\sim 95\%$.  Thus, it is likely that if such
fractal clouds exist, they would have been detected in our
observations.  A small cloud in this region, e.g. $D=1.8$,
$R_c = 110$ pc, would have 
a mass, $5\,\times\,10^7~M_\odot$
(Eqn~\EC{masscloud}) and
a velocity dispersion of 25 km s$^{-1}$
(Eqn~\EC{velcloud}).
There is another interesting region of parameter space which can be ruled
out at the 90\% 
probablity level.  This extends to much smaller clouds
($R_c < R_b$) with a higher covering factor (higher $D$).  For example,
$D = 1.9$, $R_c = 17$ pc (i.e. log($R_c/R_*$)=5.2) correponds to a
$6\,\times\,10^6~M_\odot$ cloud with a velocity dispersion of 22 km s$^{-1}$.
Over all, there is a region of parameter space with a
fractal dimension between about 1.7 and 2 and a fractal cloud 
dynamic range between about $10^5$ and $10^7$ within which there
is a greater than 50\% probability that a detection would have resulted
from our observations.

\subsubsection{Comparison with Optically Thin, Diffuse Gas}

It is fruitful to examine the more conventional assumption that the
gas is optically thin, diffuse, and that the velocity and beam filling
factors are both 1.  In this case, Fig.~4 provides upper limits to
$\tau$.  Our strongest constraint, i.e. $\tau\,<\,0.01$, then gives,
for a single channel (Eqn~\EC{nheq2}), $N(HI) \,= \,
4.7~\times~10^{16}~T_s ~{\rm cm^{-2}}$ and for $T_s\,=\,3\,K$, $N(HI)
\,= \, 1.4~\times~10^{17}~ {\rm cm^{-2}}$, illustrating the high
sensitivity of these absorption observations.  This limit is deeper
than previously measured 3$\sigma$ noise values of $N(HI) =
1~\times~10^{18}~ (T_s/f_b) {\rm cm^{-2}}$ for Ly$\alpha$ absorption
systems (Carilli et al. 1998), and is comparable to the upper limit of
$3~\times 10^{16}~T_s/f_b$ for HI absorption against more distant
cooling flow galaxies (Dwarakanath et al. 1994).  From
Eqn~\EC{surfdens}, this implies that $F_H~= 6.4~\times~10^{-6}$ for a
flattened halo (and a factor of 7 higher for a spherical halo).  Thus
no more than 0.001\% of the dark matter halo in NGC~3079 can be in the
form of cold diffuse HI.

\section{Discussion}
\label{discussion}

The detection of dark matter remains one of the most ambitious quests
in modern astrophysics.  The observations described above demonstrate
that one can search for a particular type of dark matter, cold, dense,
hydrogen clouds, by looking for absorption features in the continuum
spectrum of a background quasar.  This is, of course, reminiscent of
the rich study of Lyman-$\alpha$ clouds.  However, in the case of
NGC~3079/Q~0957+561, there is no doubt that the foreground absorber is
a galaxy, in this case one which is nearby and has been well-studied
in emission.  We are also probing, in particular, the possible
existence of halo gas far beyond the emitting HI disk which may be in
an unusual form, i.e.  cold and with a fractal structure.

The analysis indicates that for an all-cloud halo with model
parameters in a limited though interesting range (fractal dimension
$D>1.7$; dynamic range of fractal $\sim 10^5-10^7$) there was a
good chance for detection (see Fig. 8).  The
areas for improvement are obvious.  Improved resolution would extend
the region of sensitivity to lower values of $R_c/R_*$.  Better signal
to noise would bring probability levels up to where one would be able
to rule out, at say the 99\% confidence level, certain regions of
parameter space.  In addition, improved signal to noise would extend
the region of sensitivity to lower values of the fractal
dimension $D$.  Roughly speaking, we require $\left (R_b/R_* \right
)^{D-2}>3a$ where $a=(T_B-T_c)/(-T_c)$.  An improvement in signal to
noise by a factor of 10, at the same resolution, would allow us to
probe clouds with $D$ as low as $1.45$.

Since these observations were at the highest resolution possible for
connected arrays, the only practical improvements would be to search
for objects like this in our own Galaxy, to use VLBI techniques,
and/or to argue for more observing time for higher S/N.  The
difficulty in probing for objects in our own Galaxy is that the
distance to the absorber may not be well known so it may not be
possible to pin-point 
a particular absorber as a halo object.  In moving to
VLBI techniques, sensitivity is lost in comparison to connected
arrays.

One possibility might be use VLBI techniques to look for
individual Jupiter-mass cloudlets, directly, in our own Milky
Way halo.  Consider, for example, a small
``sparse" cloud with $D\,=\,1.3$ and $\left(R_c/R_*\right)\,=\,10^3$
($R_*\,=\, 23$ au, $R_c\,=\,0.1$ pc), at a distance of 30 kpc which is
distant enough to be far from Galactic sources of heating.  An
individual cloudlet will subtend an angle of 1.5 milliarcseconds while
the cloud will subtend an angle, 1.5 arcseconds.  The mean distance
between clouds will be ${\left({R_c}^2/{\cal
N}_c\right)}^{1/2}\,=\,1\,\times\,10^{-3}$ pc (see
Eqn~\EC{masscloud}), or 8 milliarcseconds.  These parameters are
favourable for multiple detections over a VLBI field of view against a
strong background source.  If the cloudlets are indeed optically thick
at 3 $K$, the absorption signature should be obvious.

Indeed, VLBI HI absorption measurements have already detected a
population of objects in the Galaxy with size scales of order 10s of
au, as summarized and discussed by Heiles (1997).  These tiny HI
clouds, referred to by Heiles as ``tiny scale atomic structure" (TSAS)
have densities of $\sim$ 10$^4$ cm$^{-3}$.  This is in the
range of $10^3\,\to\,10^5$ cm$^{-3}$ which we have estimated for the
HI fraction of PC cloudlets (\S\ref{composition}).  TSAS features are
relatively nearby as indicated by velocities, $|v|\,<\,20$ km s$^{-1}$
(Diamond et al. 1989, Davis et al. 1996) or by absorption against
pulsars which are within $\sim$ 3 kpc (Frail et al. 1994).
Consequently, they are near Galactic sources of heating (temperatures
are often assumed to be $\sim\,50\,K$).  Given the sizes and densities
of these objects, the possibility is raised that they are PC cloudlets
which have already been accreted onto the Milky Way.  A clear
detection of CO associated with the HI absorbing systems would rule
out this possibility.  

To resolve the problem of unknown distance
and argue that such objects are in the Galactic halo, 
a population of such objects
would have to be detected at high velocities over a number of lines
of sight.

\section{Summary}
\label{summary}

Pfenniger et al. (1994) and Pfenniger \& Combes (1994) have proposed a
new model of dark matter consisting of cold clouds, radiatively
coupled to the $3~K$ background, in an extended disk around galaxies.
This gas is proposed to consist of virialized elementary cloudlets and
to be of primordial composition except that at their high densities
($\sim\,6\,\times\,10^9$ cm$^{-3}$), gas phase reactions can convert
most of the HI to $H_2$.  The cloudlets are arranged in a hierarchical
fractal structure into larger clouds, the fractal structure being
essential to ensuring that collisions do not dissipate the cloud and,
in combination with low temperatures, that stars do not form.

In this work, we have shown that, at the temperatures and densities
envisioned, a small fraction of HI will remain in the cloudlets and
will likely be optically thick.  Thus, their detectability depends on
their temperature and the filling factor of the resolution element,
rather than quantity.  At a temperature of $3~K$, the clouds cannot be
seen in emission against the microwave background, but HI is
potentially observable in absorption against a background source.
This means that HI can be used as a probe of PC-type fractal clouds
and opens a new window for exploring halo dark matter which may be in
this form.

Using the VLA, we have carried out the first observational test of the
PC model by taking advantage of a fortuitous alignment of a strong
background quasar at a projected separation of 64 kpc from the center
of the nearby galaxy, NGC~3079.  The quasar is aligned along an
extension of the galaxy's major axis.  By tuning to the velocity of
the foreground galaxy, we made 21 cm absorption measurements against
the quasar, obtaining high spatial resolution (1.2$\arcsec$, or 28
beams across the background continuum) and high velocity resolution
(2.6 km s$^{-1}$).  We do not detect any HI absorption to a (minimum)
3$\sigma$ upper limit $3\,\Delta\,T_{B_v}/{(-T_c)}$ = 0.01.

Our observational limits can be related to the filling factor of
fractal clouds in a beam/velocity resolution element and we have
considered, in detail, how the observations might constrain parameter
space, defined by the quantities, $R_c/R_*$, $D$, and $F_H$, i.e. the
dynamic range of the fractal structure, the fractal dimension, and
the fraction of dark matter contained in such clouds,
respectively.  We find that
much of parameter space cannot be ruled out by our observations.  
However, there is
 a region of parameter space (i.e., $1.7\la D\la 2$ and
$30\,{\rm pc}\la R_c\la 3\,{\rm kpc}$) within which there
is a reasonable probability of
detection.  In the region $1.75\la D\la 1.85$ and
$100\,{\rm pc}\la R_c\la 3\,{\rm kpc}$,
the probability of detection is as high as
 $\sim 95\%$
(see Fig. 8).  Thus, it is unlikely that clouds
with these parameters exist.  

It is interesting that cold diffuse (optically thin) HI can be ruled
out to a limit of 0.001\% of the dark matter.  In contrast, 
congregating the HI into optically thick fractal clouds is a very
efficient way of hiding dark matter.  Our analysis, applicable to any
fractal clouds containing optically thick HI, is the first step in
probing dark matter in this form.  We outline possible ways of
improving the observations in the future.

\acknowledgments{ The authors wish to thank T. Abel, M. Mandy,
R. Henriksen, N. Pelavas, and M. Walker for useful discussions.  We
have made used of the NASA/IPAC Extragalactic Database (NED). This
work is supported by the Natural Sciences and Engineering Research
Council of Canada}

\newpage

\appendix{\bf Appendix}

In this appendix we calculate $P(\nb|\ne)$, the probability that at
least one channel in a given beam will contain $\ne$ or more clouds
given $\nb$ clouds in this beam.  The clouds are characterized by an
internal velocity dispersion, $v_c$ (assumed to be the same for all
clouds) and a bulk velocity, $\bar v$ which is different for each of
the clouds.  For a given channel to contain contributions from two
clouds, we require that their bulk velocities be separated by less
than ${\rm max}(v_c-\Delta v,\Delta v)$.  That is, for $v_c>2\Delta
v$, we require that the overlap of the clouds in velocity space should
be greater than the width of a single channel whereas for $v_c<2\Delta
v$ we require that the separation of the clouds in velocity space be
less than $\Delta v$.  To continue, we need to know the distribution
function for $\bar v$.  In the absence of a detailed model for the
cloud population, we assume that the $\bar v$ are equally likely to
reside in any part of the $163\,{\rm km\,s^{-1}}$ velocity range
covered by our observations.  The probability that two particular
clouds will contribute to a single channel is therefore $p={\rm
max}(v_c-\Delta v,\Delta v)/163\,{\rm km\,s^{-1}}$.  Put another way,
we can think of there being $L\equiv 1/p$ bins in velocity space and
we are asking for the probability that a certain number of clouds will
appear in a single bin.

In general, if there are $\nb$ clouds in the beam, then the 
the probability that $l_0$ bins contain $0$ clouds, 
$l_1$ bins contain one cloud, etc is given by

\be{plm}
P(\nb,l_m,p)~=~p^\nb\,\frac{L!}{\Pi_m l_m!}\,\frac{\nb!}{\Pi_m \left (m!
\right )^{l_m}}\eqnum{A1}
\ee

\noindent The first factor is simply the probability of having a
particular combination of $\nb$ clouds in $L$ bins.  The second
factor handles the fact that we don't care which bins contain the
different numbers of clouds.  The final factor handles the fact that
we don't care which clouds go into which bin.  The desired probability
is then given by

\be{pnbne1}
P(\nb|\ne)=\sum P(\nb,l_m,p)\eqnum{A2}
\ee

\noindent where the sum is over all combinations of $l_m$ consistent
with the constraints that $\sum_m l_m=1/p$, $\sum_m l_m m = \nb$ and
$l_M\ge 1$.

To illustrate how the calculation will go, we consider a few simple
cases.  Clearly $P(\nb|1)=1$ since the probability of having
at least one channel with one or more clouds is unity provided
$\nb\ge 1$.  As discussed above, and in agreement
with Eq\,A1 and A2, $P(2|2)=p$.  For $P(4|2)$,
there will be four contributions to Eq.\,A2 corresponding
to $(l_0,l_1,l_2,l_3,l_4)=(1,2,1,0,0),(2,1,0,1,0),(3,0,0,0,1),
(2,0,2,0,0)$.  For the special case where $l=1/p=4$ we find
$P(4|2)=29/32$.  This case is easy to check since the probability
of having two or more clouds in one of the four channels is 
simply $1-({\rm probability~of~having~one~cloud~in~each~channel})
=1-4!/4^4=29/32$.

\newpage

\figcaption[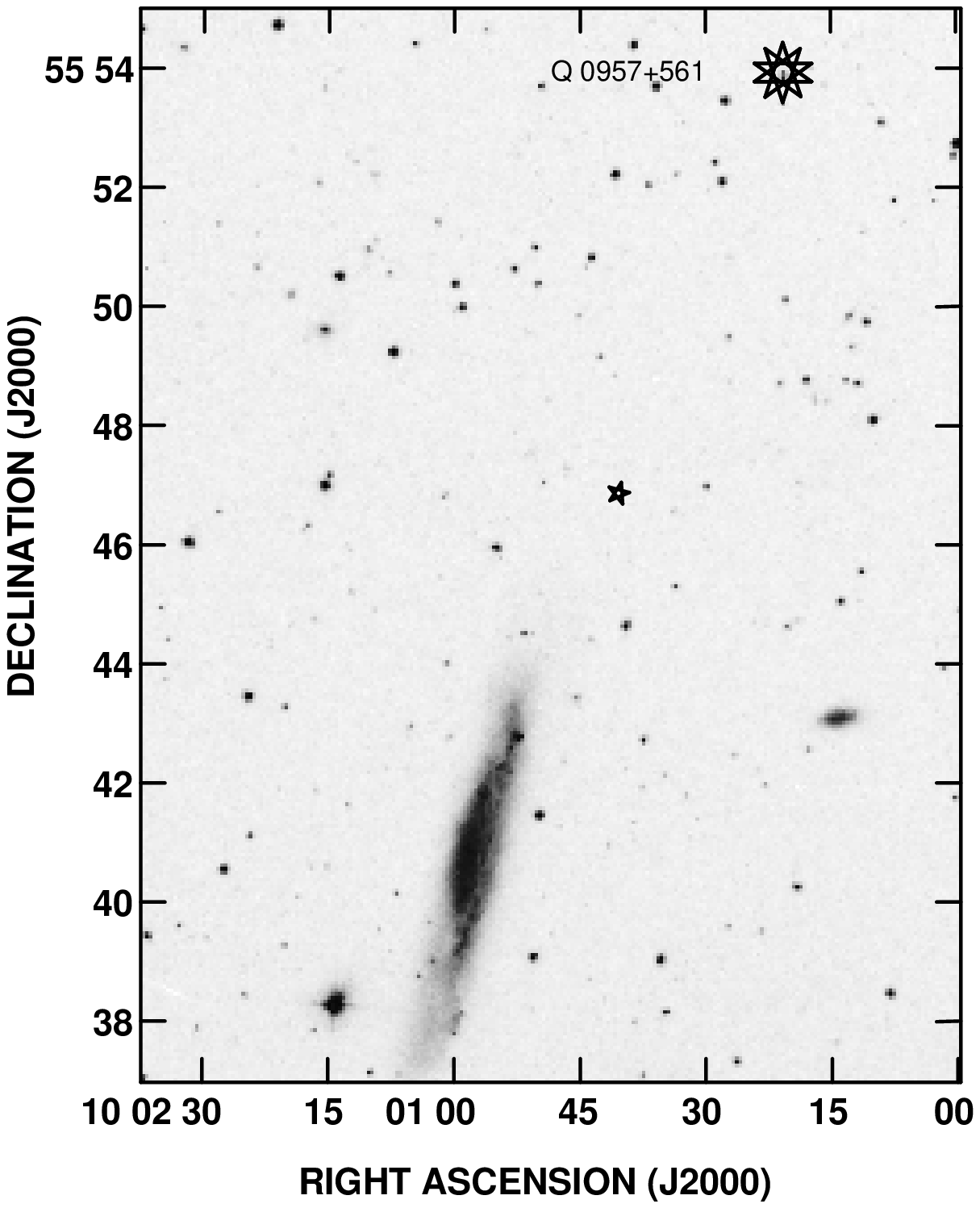]{ Digitized Sky Survey optical image of the
foreground galaxy, NGC~3079, and the background quasar, Q~0957+561
(large star).  The small star marks the approximate boundary of the HI
emission associated with NGC~3079 as measured by Irwin et al. (1987).
\label{fig1}}

\figcaption[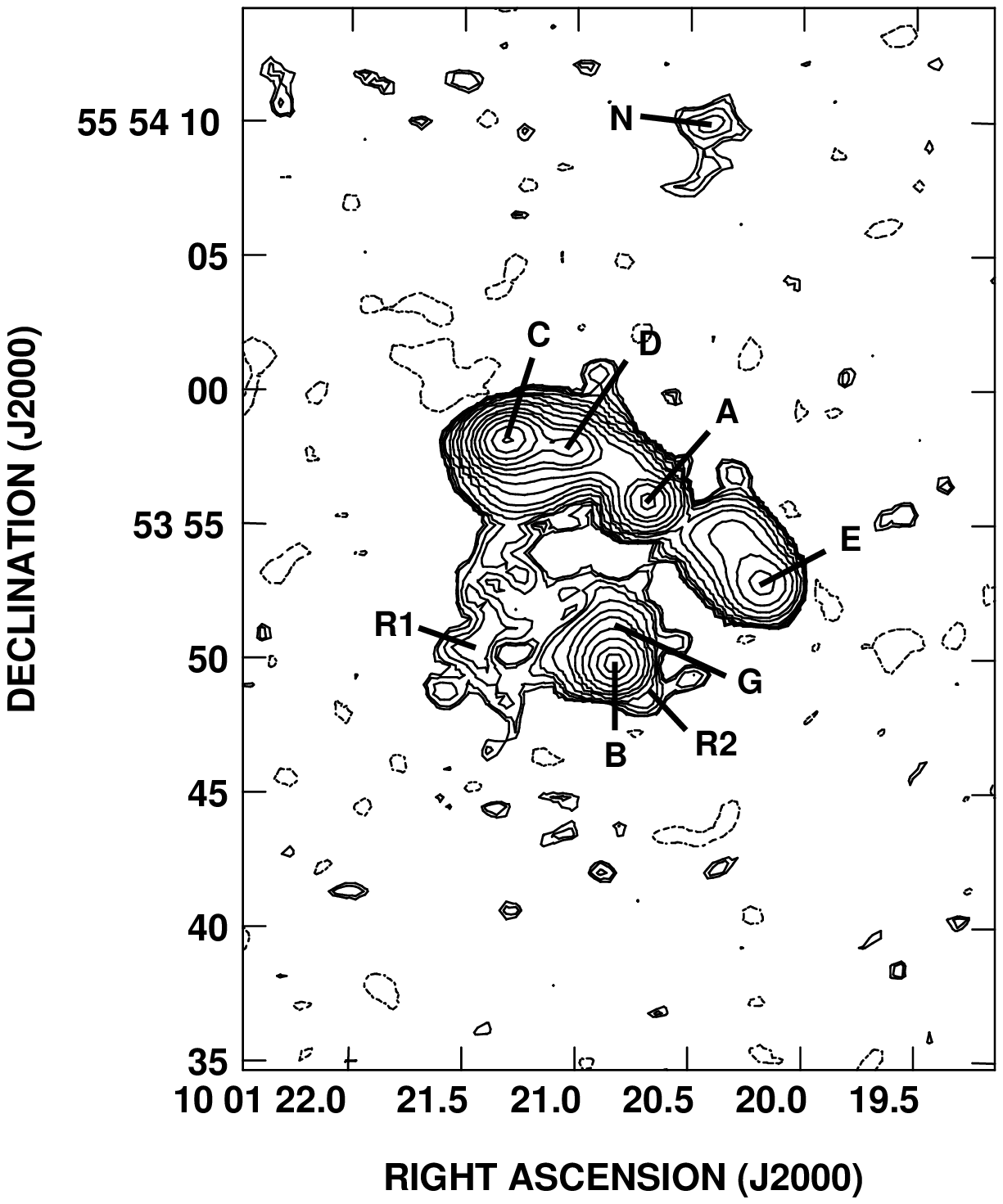]{Uniformly weighted continuum map of
Q0957+561. Contours are at -0.26, 0.26 (2$\sigma$), 0.3, 0.4, 0.5,
0.8, 1.5, 2.5, 5, 10, 20, 30, 50, 75, 120, and 170 mJy beam$^{-1}$.
The peak intensity is 174.5 mJy beam$^{-1}$ and the beam size is
1.31$\sec$ $\times$ 1.15$\sec$ at position angle, -87.31$\deg$.
Labelling follows the conventions of Greenfield et al. (1985) and
Avruch et al. (1997).  The lensing galaxy is associated with radio
source, G.\label{f2}}

\figcaption[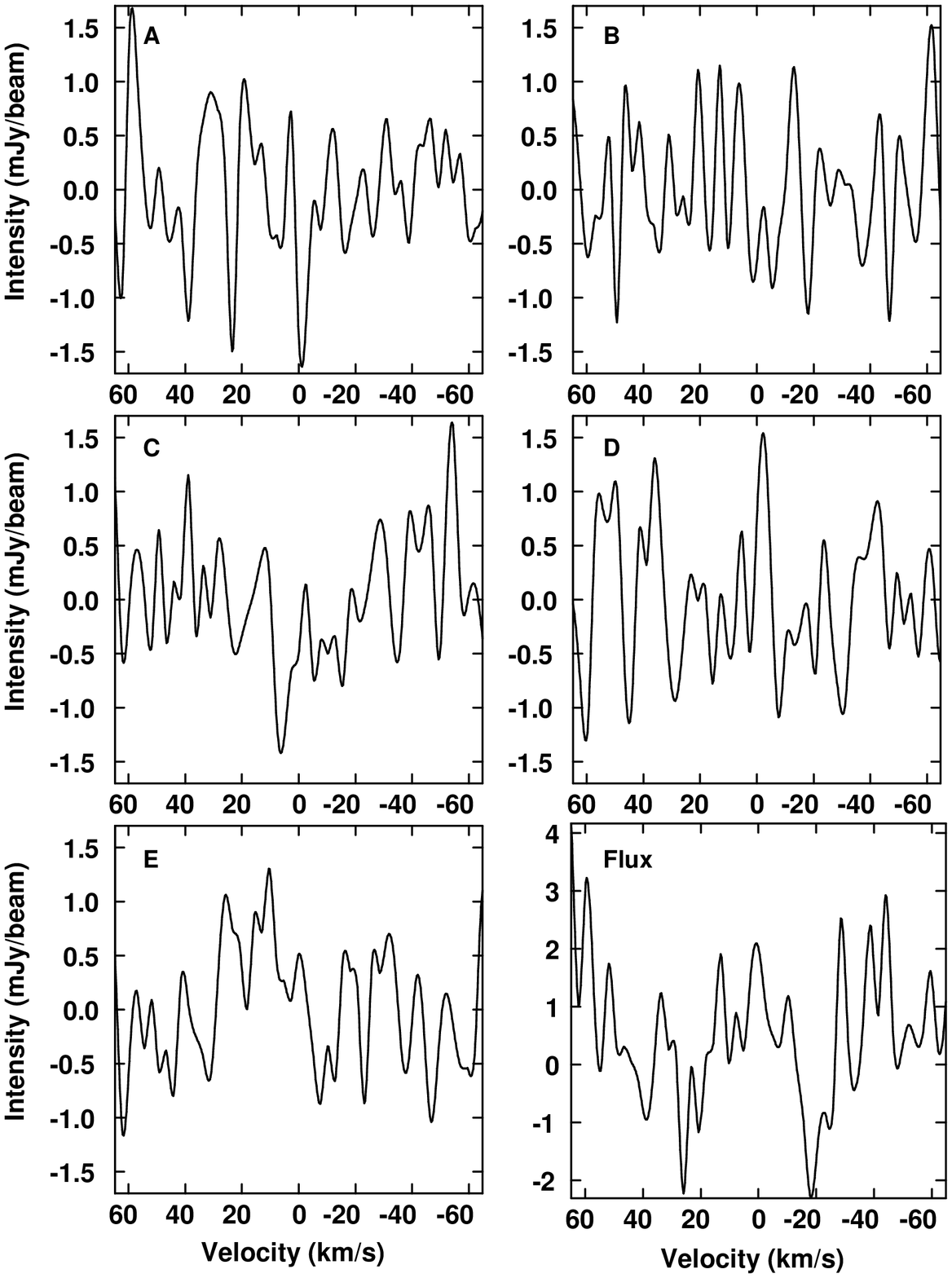] {Spectra through the continuum-subtracted
channel maps at positions corresponding to the peak of components A,
B, C, D, and E (see Fig. 2).  The rms map noise per channel is 0.58
$\pm$ 0.01 mJy beam$^{-1}$ and the velocity is with respect to 974 km
s$^{-1}$.  The last panel shows the integrated flux (mJy) measured
over a region within which the continuum emission is greater than
10$\sigma$.
\label{f3}}

\figcaption[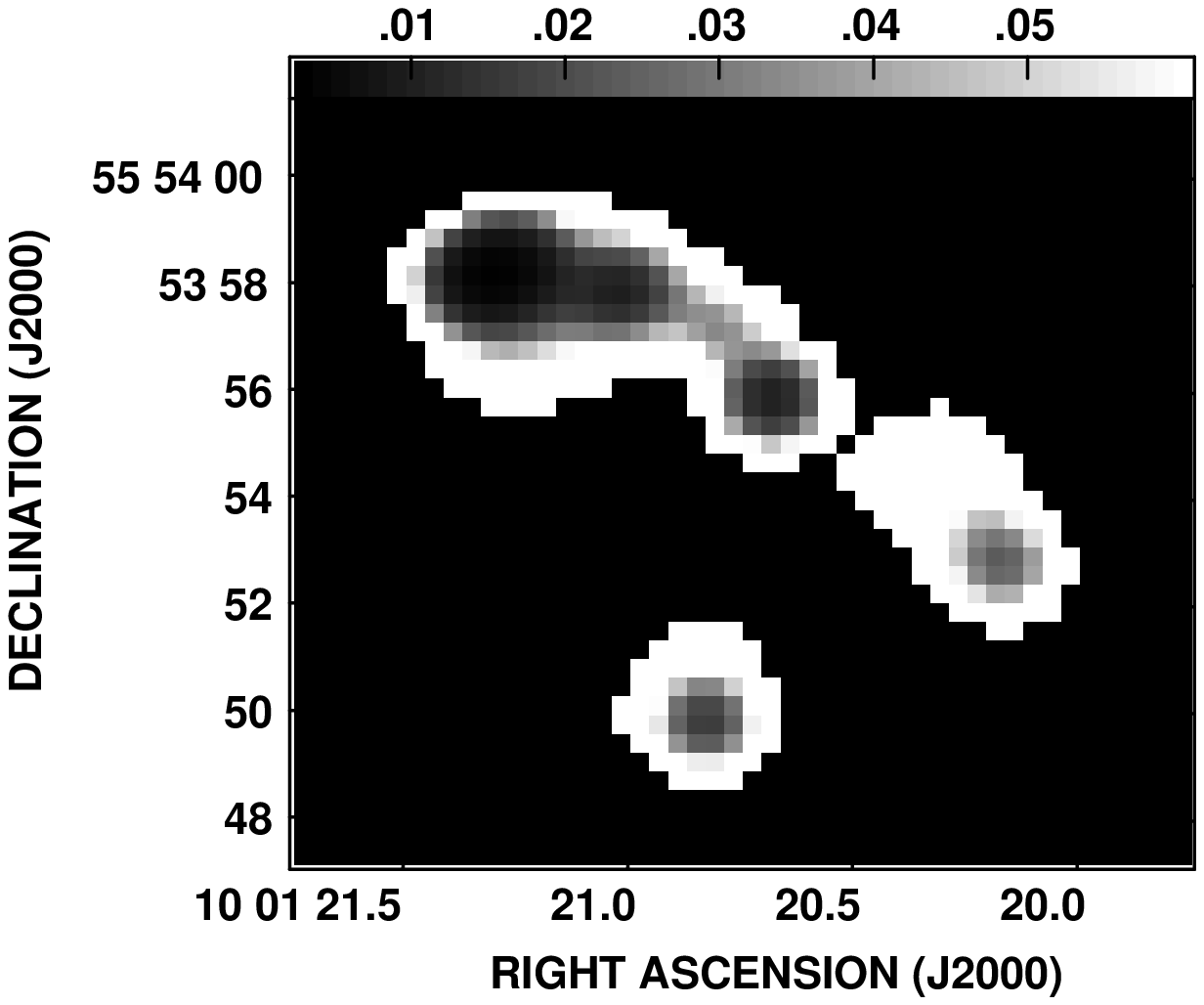]{Map of upper limits on $f_b\,f_v$ (if the
gas is optically thick) or on $\tau ~f_b~f_v$ (if the gas is
optically thin) for a single channel, assuming that the upper
limit on $\Delta {T_B}$ is given by 1 $\times$ the rms map noise.
The map values range from a minimum of 0.00332 to a maximum of 0.385
and the greyscale ranges from 0.003 (black) to 0.06 (white).  External
black areas have been blanked.
\label{f4}}

\figcaption[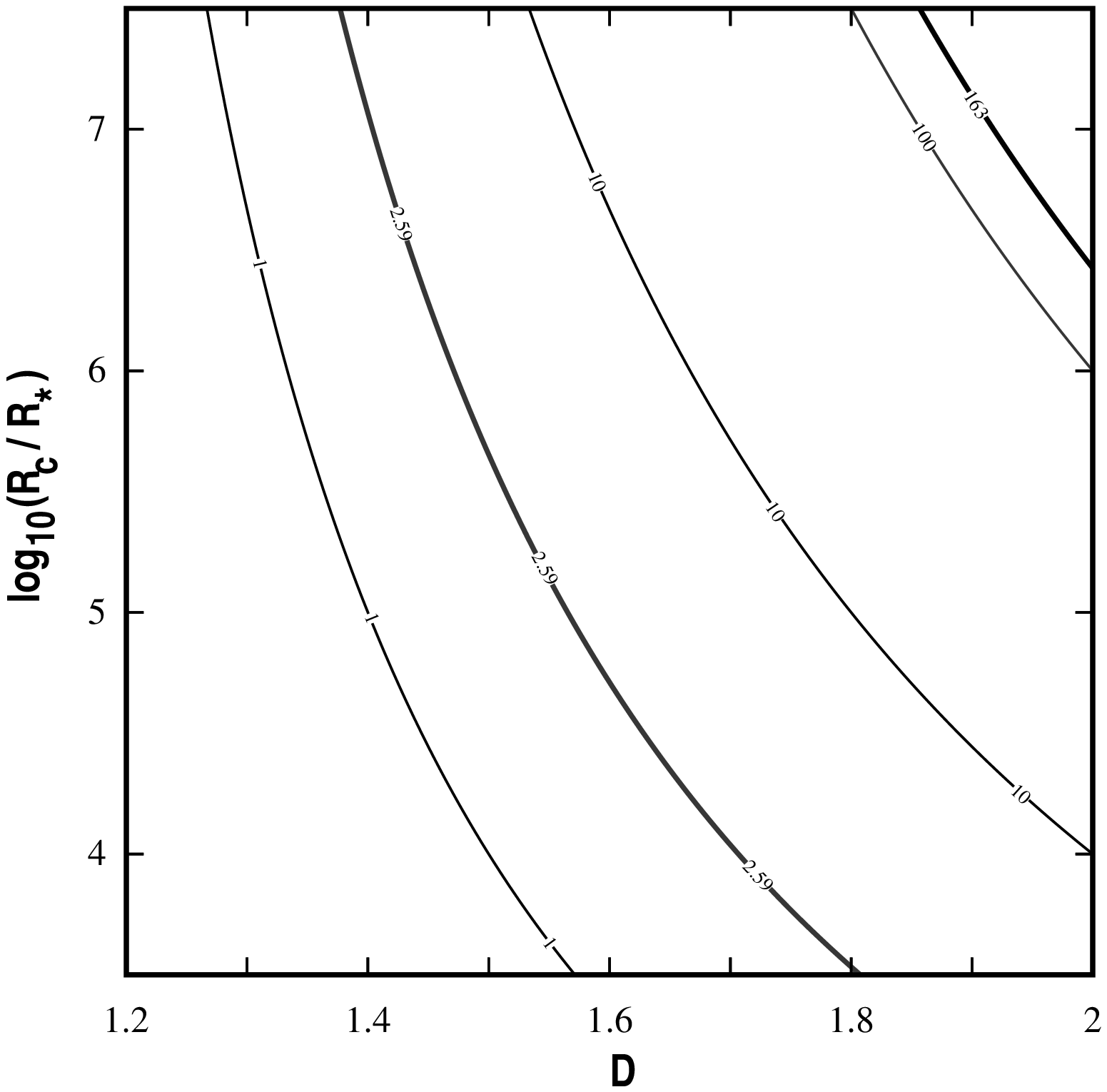]{Contour plot of the cloud velocity dispersion
$v_c$ (in units of ${\rm km\,s^{-1}}$)
as a function of model parameters $R_c/R_*$ and $D$.  Contours are
as labelled.  The $2.59$ and $163$ contours correspond to $v_c=\Delta v$
and $v_c$ equal to the velocity range of the observations respectively.
\label{f5}}

\figcaption[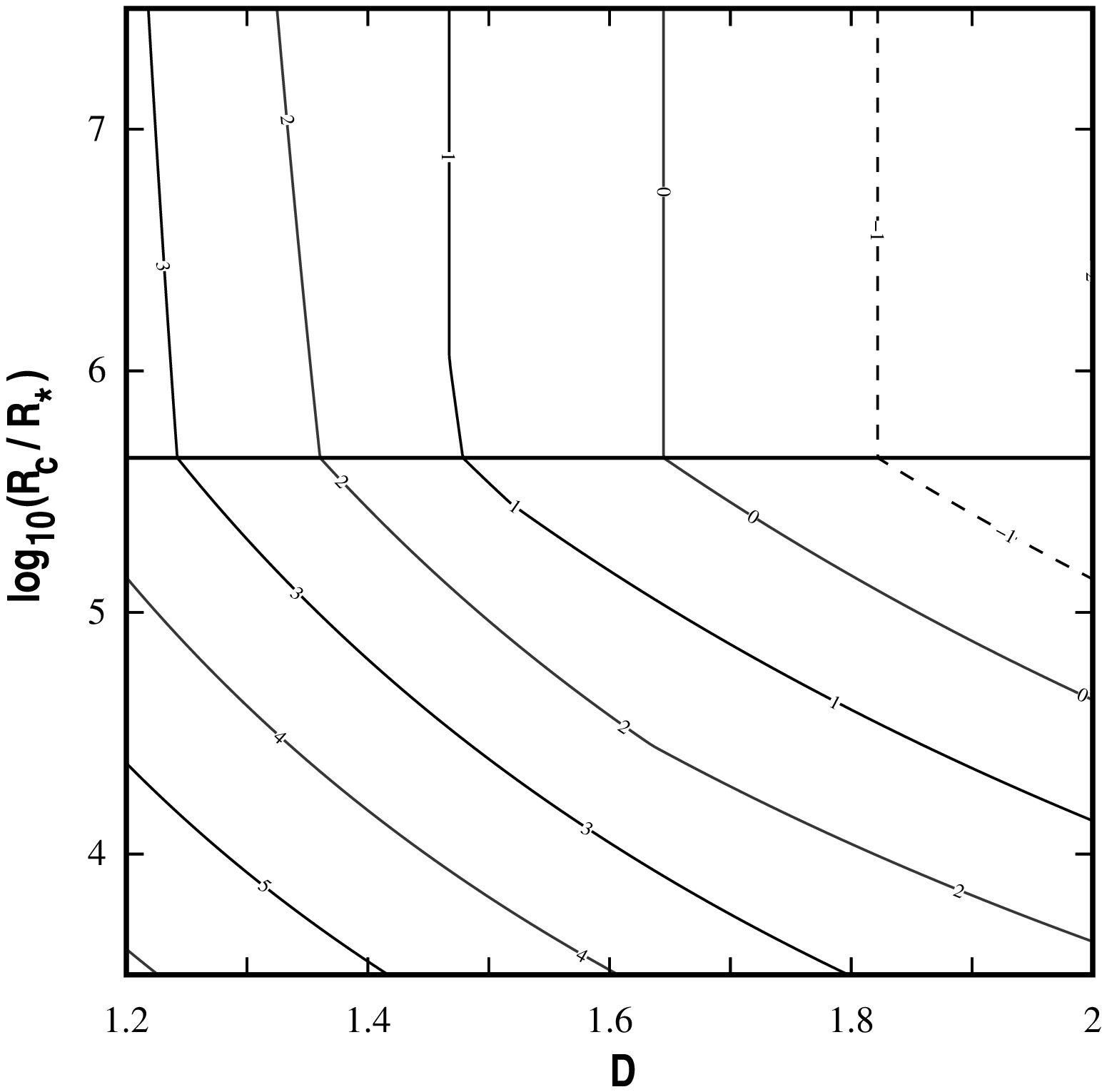]{Contour plot of $\nm$.  Contours are 
at equally spaced intervals in $log_{10}(\nm)$.  The horizontal
line corresponds to the 
beam radius.   
\label{f6}}

\figcaption[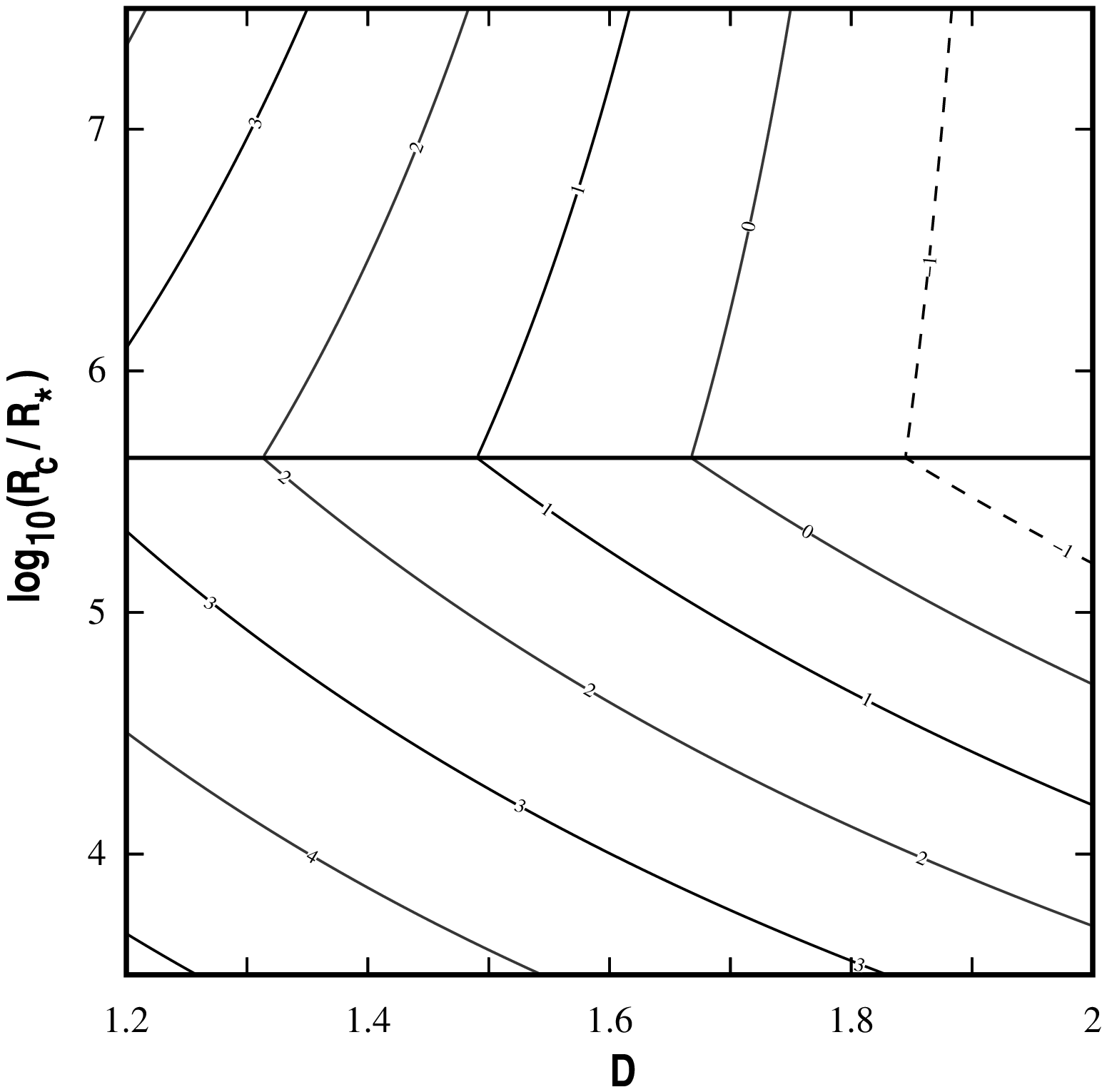]{Contour plot of $\enb$.  Contours are 
at equally spaced intervals in $log_{10}(\enb)$.  We have assumed
$F_H=1$.
\label{f7}}

\figcaption[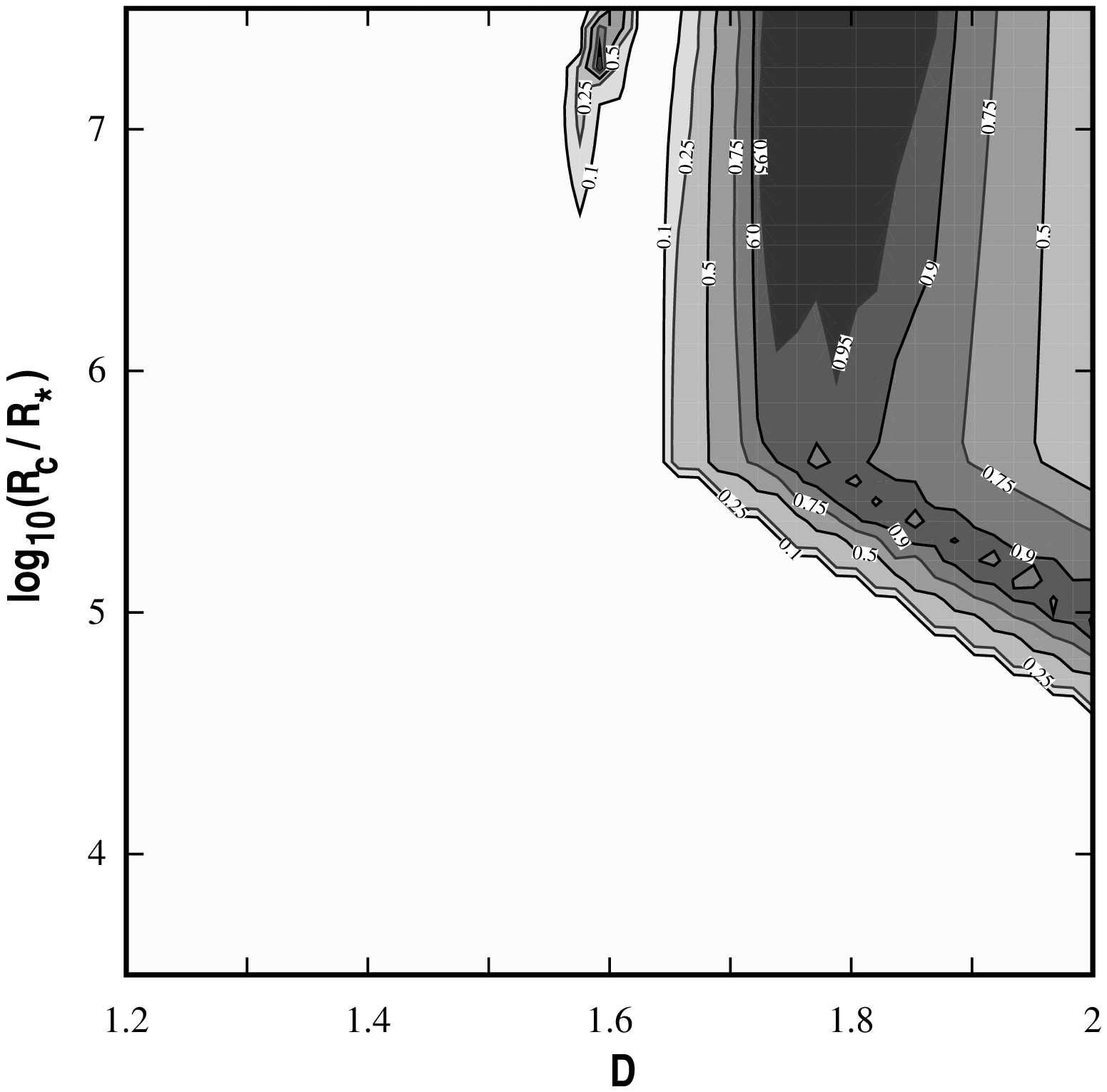]{Contour plot of the probability ${\cal P}$
for detecting a cloud as a function of the model parameters 
$R_c/R_*$ and $D$.  For this figure, we assume $F_H=1$.  Probability
contours are as labelled.
\label{f8}}

\clearpage

\begin{figure}
\begin{center}
\psfig{file=figure_1.ps}
\label{figure_1}
\end{center}
\end{figure}

\begin{figure}
\begin{center}
\centerline{\hbox{
\psfig{file=figure_2.ps}
}}
\label{figure_2}
\end{center}
\end{figure}

\begin{figure}
\begin{center}
\psfig{file=figure_3.ps}
\label{figure_3}
\end{center}
\end{figure}

\begin{figure}
\begin{center}
\psfig{file=figure_4.ps}
\label{figure_4}
\end{center}
\end{figure}

\begin{figure}
\begin{center}
\psfig{file=figure_5.ps}
\label{figure_5}
\end{center}
\end{figure}

\begin{figure}
\begin{center}
\psfig{file=figure_6.ps}
\label{figure_6}
\end{center}
\end{figure}

\begin{figure}
\begin{center}
\psfig{file=figure_7.ps}
\label{figure_7}
\end{center}
\end{figure}

\begin{figure}
\begin{center}
\psfig{file=figure_8.ps}
\label{figure_8}
\end{center}
\end{figure}


\newpage

\begin{thebibliography}{}

\bibitem[]{alc97} Alcock et al.\,1997a, \apj, 486, 697
\bibitem[]{al97} Alcock et al.\,1997b, \apj, 497, L11
\bibitem[]{ash92} Ashman, K. M. 1992, \pasp, 104, 1109
\bibitem[]{avr97} Avruch, I. M., Cohen, A. S., Leh\'ar, J., Conner, S. R.,
Haarsma, D. B., \& Burke, B. F. 1997, \apj (letters), 488, L121
\bibitem[]{bla97} Bland-Hawthorn, J., Freeman, K., \& Quinn, P. J. 1997,
\apj, 490, 143
\bibitem[1994]{car94} Carr, B. 1994, \araa 32, 531
\bibitem[] {car98} Carilli, C. L., Menten, K. M., Reid, M. J., Rupen, M. P.,
\& Yung, Min Su 1998, \apj, 494, 175
\bibitem[]{coh83}Cohen, N., and Westberg, K. R. 1983, J. Phys. Chem. Ref. Data,
12, 560
\bibitem[]{cop95} Copi, C. J., Schramm, D. N., \& Turner, M. S. 1995,
Science, 267, 192 
\bibitem[]{cor93} Corbelli, E., \& Salpeter, E. E. 1993, \apj, 419, 94
\bibitem[]{corb89} Cornwell, T., \& Braun, R. 1989
 in ASP Conf. Series 6, Synthesis
Imaging in Radio Astronomy, ed. R. A. Perley, F. R. Schwab, \& A. H.
Bridle, p. 167
\bibitem[]{corf89} Cornwell, T., \& Fomalont, E. B. 1989
 in ASP Conf. Series 6, Synthesis
Imaging in Radio Astronomy, ed. R. A. Perley, F. R. Schwab, \& A. H.
Bridle, p. 185
\bibitem[]{dav96} Davis, R. J., Diamond, P. J., \& Goss, W. M. 1996,
\mnras, 283, 1105
\bibitem[Deguchi \& Watson 1985]{deg85}
 Deguchi, S., \& Watson, W. D. 1985, \apj, 290, 578
\bibitem[1989]{dia89} Diamond, P. J., Goss, W. M., Romney, J. D., Booth, R. S.,
Kalberla, P. M. W., \& Mebold, U. 
 1989, \apj, 347, 302
\bibitem[Dove \& Shull 1994]{dov94}
 Dove, J. B., \& Shull, J. M. 1994, \apj, 423, 196
\bibitem[]{dwa94}
Dwarakanath, K. S., van Gorkom, J. H., \& Owen, F. N. 1994,
\apj, 432, 469
\bibitem[]{eke89} Ekers, R. D. 1989, in ASP Conf. Series 6, Synthesis
Imaging in Radio Astronomy, ed. R. A. Perley, F. R. Schwab, \& A. H.
Bridle, p. 199
\bibitem[] Elmegreen, B. 1996, \apj, 471, 816
\bibitem[Filippenko \& Sargent 1992]{fil92}
 Filippenko, A. V., \& Sargent, W. L. W. 1992, \aj, 103, 28
\bibitem[]{fra94} Frail, D. A., Weisberg, J. M., Cordes, J. M., \&
Mathers, C. 1994 \apj, 436, 144
\bibitem[]{gal98} Galli, D., \& Palla, F. 1998, \aa, 335, 403
\bibitem[]{gre85} Greenfield, P. E., Roberts, D. H., \& Burke, B. F.
1985, \apj, 293, 370
\bibitem[]{haa96}  Haardt, F., \& Madau, P. 1996, \apj, 461, 20
\bibitem[]{hei97} Heiles, C. 1997, \apj, 481, 193
\bibitem[Henriksen \& Widrow]{hen95} Henriksen, R. N., \& Widrow, L. W.
1995, \apj, 441, 70
\bibitem[Irwin 1995]{irw95} Irwin, J. A. 1995, \pasp 107, 715
\bibitem[Irwin \& Seaquist 1988]{irw88}
 Irwin, J. A., \& Seaquist, E. R. 1988, \apj, 335, 658
\bibitem[Irwin \& Seaquist 1990]{irw90}
Irwin, J. A., \& Seaquist, E. R. 1990, \apj, 353, 469
\bibitem[Irwin \& Seaquist 1991]{irw91} 
Irwin, J. A., \& Seaquist, E. R. 1991, \apj, 371, 111
\bibitem[]{irw87} Irwin, J. A., Seaquist, E. R., Taylor, A. R., \& Duric, N.
1987, \apj, 313, L91
\bibitem[]{irw99}  Irwin, J. A., Widrow, L. M., \& English, J. 1999,
PASA, in press
\bibitem[]{ken94} Kennicutt, R. C. Jr., Tamblyn, P., \&
Congdon, C. W. 1994, \apj, 435, 22
\bibitem[Maloney 1993]{mal93} Maloney, P. 1993, \apj 414, 41
\bibitem[1983]{pal83} Palla, F., Salpeter, E. E., \& Stahler, S. W. 1983,
\apj, 271, 632
\bibitem[Pfenniger \& Combes 1994]{pc94}
 Pfenniger, D., \& Combes, F. 1994, \aap, 285, 94 (PC)
\bibitem[Pfenniger et al. 1994]{pcm94}
 Pfenniger, D., Combes, F., \& Martinet, L. 1994, \aap, 285, 79 (PCM)
\bibitem[Rubin, Ford, \& Thonnard 1980]{rub80}
Rubin, V. C., Ford, W. K., \& Thonnard, N. 1980, \apj, 238, 471
\bibitem[van Gorkom 1993]{vgo93}
van Gorkom, J. H. 1993, in The Environment and Evolution of
Galaxies, ed. J. M. Shull and H. A. Thronson, Jr. (Dordrecht,
Kluwer), p. 345
\bibitem[]{sch98} Schmidt, R., \& Wambsganss, J. 1998, \aap, 335, 379
\bibitem[]{ste99} Steigman, G., Hata, N., \& Felten, J. E. 1999, \apj,
510, 564
\bibitem[]{swa89} Sramek, R. A., \& Schwab, F. R. 1989
 in ASP Conf. Series 6, Synthesis
Imaging in Radio Astronomy, ed. R. A. Perley, F. R. Schwab, \& A. H.
Bridle, p. 185
\bibitem[]{sta99} Stanimirovic, S., Staveley-Smith, L., Dickey, J. M.,
Sault, R. J., \& Snowden, S. L. 1999, \mnras, 302, 417
\bibitem[]{vog94} Vogelaar, M. G. R., \& Wakker, B. P. 1994, \aa, 291, 557
\bibitem[Walsh et al. 1979]{wal79}
 Walsh, D., Carswell, R. F., \& Weymann, R. J. 1979, Nature,
279, 381
\bibitem[]{wes99} Westpfahl, D. J., Coleman, P. H., Alexander, J.,
\& Tongue, T. 1999, \aj, 117, 868

\end{thebibliography}
\end{document}